\documentstyle[prd,preprint,aps]{revtex}

\begin{document}
\draft
\preprint{January-95}
\title{Properties of vector mesons at finite temperature\\
       $-$effective lagrangian approach$-$}
\bigskip
\author{Chungsik Song}
\address{ Cyclotron Institute, Texas A\&M University,
          College Station, TX 77843, USA }
\maketitle
\begin{abstract}
The properties of $\rho$-mesons at finite temperature ($T$)
are examined with an effective chiral lagrangian
in which vector and axial-vector mesons are included
as massive Yang-Mills fields of the chiral symmetry.
It is shown that, at $T^2$ order, the effective mass is not changed
but only the mixing effect in
vector and axial-vector correlator appears.

\end{abstract}
\vspace{0.5cm}

\section{Introduction}

It is expected
that spontaneously broken chiral symmetry is restored
in hadronic matter at very high temperature \cite{chiral},
which has been shown for QCD on lattice \cite{lattice}.
The restoration of the chiral symmetry is characterized by
melting out of the quark condensate which is known as the order parameter
of the chiral phase transition. The vanishing of the quark condensate
with temperature also has been shown from some model calculations
\cite{gl}.

We are interest in the phenomena which arise prior to the chiral
transition in hot hadronic matter. Even though the quark condensate is not
directly observable the change of the condensate at finite temperature
would affect on the properties of hadrons.
It is desirable to have a direct connection between the properties
of hadrons and those of ground state
in order to study the chiral transition in hot matter.
Of particular interest are the vector meson properties at finite temperature
since model calculations show definite relation between chiral symmetry
restoration at finite temperature and vector meson mass \cite{pisa}.
Moreover, the change in the vector meson mass
can be observed from the shift of the vector meson peaks in dilepton spectrum
from hot matter.

The vector meson properties and dilepton spectrum at finite temperature
have been studied in various ways
\cite{pisa,sum,dey,hatsuda,eletsky,gale,song,ko,slee}.
Recently, it was realized by Dey et.~al. \cite{dey} that
in the lowest order of
$\epsilon=T^2/6f_\pi^2$, where the pion decay constant $f_\pi=93$ MeV,
there is no change
in the vector meson mass and only mixing between vector and axial-vector
correlator takes place. Since the isospin mixing effect is obtained based only
on the PCAC and current algebra, it has to be satisfied in the low
temperature limits of any model calculations as long as the model
calculations preserve the same symmetry properties.
This constraint is indeed satisfied in the low temperature limit
of the calculation using QCD sum rule \cite{hatsuda,eletsky}
and with the effective chiral lagrangian \cite{slee}.

In this paper the isospin mixing effect is obtained from an
effective lagrangian approach in which
vector and axial-vector mesons are included
as massive Yang-Mills fields of the chiral group and the photon
fields are introduced via vector meson dominance assumption.
The result indicates that
the effective masses of vector mesons are not modified in the
leading order of temperature, ${\cal O}(T^2)$.
In section 3, we introduce an extra non-minimal coupling term
to reproduce the
experimental values of axial-vector meson mass and width.
With the new term we cannot get the exact mixing effect
in vector and axial-vector correlator at finite temperature
but the effective masses of vector mesons
are still not changed much at the leading $T^2$ order.

\section{Isospin mixing at finite temperature}

We consider an effective chiral lagrangian with vector and axial-vector meson
fields which are introduced as massive Yang-Mills fields \cite{mym};
\begin{eqnarray}
{\cal L} &=&  {1\over4}f_\pi^2{\rm Tr}[D_\mu U D^\mu U^\dagger]\cr
         & & -{1\over2}{\rm
Tr}[F^L_{\mu\nu}F^{L\mu\nu}+F^R_{\mu\nu}F^{R\mu\nu}]
             +m_0^2{\rm Tr}[A^L_\mu A^{L\mu}+A^R_\mu A^{R\mu}]\cr
         & & -i\xi {\rm Tr}[D_\mu U D_\nu U^\dagger F^{L\mu\nu}
                          +D_\mu U^\dagger D_\nu U F^{R\mu\nu}],
\end{eqnarray}
where $U$ is related to the pseudoscalar fields $\phi$ by
\begin{equation}
U=exp\left[{i\sqrt{2}\over f_\pi}\phi\right],
\quad\phi=\sum_{a=1}^3\phi_a{\tau_a\over\sqrt{2}},
\end{equation}
and $A^L_\mu (A^R_\mu)$ are left(right)-handed vector fields.
The covariant derivative acting on $U$ is given by
\begin{equation}
D_\mu U=\partial_\mu U-igA^L_\mu U-igUA^R_\mu,
\end{equation}
and $F^L_{\mu\nu}\,(F^R_{\mu\nu})$ is the field tensor
of left(right)-handed vector fields. The $A^L_\mu$ and $A^R_\mu$
can be written in terms of vector ($V_\mu$) and
axial-vector fields ($A_\mu$) as
\begin{equation}
A_\mu^L={1\over2}(V_\mu-A_\mu), \quad A_\mu^R={1\over2}(V_\mu+A_\mu).
\end{equation}

The lagrangian can be diagonalized in the conventional way by the definitions
\begin{eqnarray}
A_\mu &\to& A_\mu+{gf_\pi\over \sqrt{2}m_0^2}
            \left(\partial_\mu\phi-i{g\over2}[V_\mu,\phi]\right),\cr
\phi  &\to& Z^{-1}\phi, \cr
f_\pi &\to& Z^{-1}f_\pi.
\end{eqnarray}
In terms of new fields we find
\begin{eqnarray}
{\cal L}^{(2)} &=& {1\over2}{\rm Tr}
                   \left(\partial_\mu\phi-i{g\over2}[V_\mu,\phi]\right)^2
         -{1\over4}{\rm Tr}[F^V_{\mu\nu}F^{V\mu\nu}+F^A_{\mu\nu}F^{A\mu\nu}]\cr
           & &  +{1\over2}m_\rho^2{\rm Tr}V_\mu^2
                +{1\over2}m_a^2{\rm Tr}A_\mu^2,
\end{eqnarray}
where we use
\begin{equation}
Z^2 = \left[1-{g^2f_\pi^2\over 2m_\rho^2}\right]={m_\rho^2\over m_a^2},
\end{equation}
and the vector and axial-vector meson mass are given by
\begin{equation}
m_\rho^2=m_0^2,\qquad m_a^2=m_0^2/Z^2.
\end{equation}
When we choose $Z^2={1\over2}$
we have the KSRF relation, $m_\rho^2=g^2f_\pi^2$, and Weinberg mass relation,
$m_a^2=2m_\rho^2$.

One can calculate the width of $\rho$-mesons from the given lagrangian as
\begin{equation}
\Gamma(\rho\to\pi\pi)={1\over6\pi m_\rho^2}\vert q_\pi\vert^3g_{\rho\pi\pi}^2
\end{equation}
with
\begin{equation}
g_{\rho\pi\pi}={g\over4\sqrt{2}}(3+2g\xi)
\end{equation}
To satisfy the Universality  of vector meson coupling
$g_{\rho\pi\pi}=g/\sqrt{2}$ \footnote{There is a factor $1/\sqrt{2}$
because $V_\mu={\tau^a\over\sqrt{2}}\rho_\mu^a$,
where $a$ is the isospin index},
we choose $2g\xi=1$.
The coupling constant $g$ can be determined from the experimental value of the
$\rho$-width, $\Gamma^{exp}(\rho\to\pi\pi)=150$ MeV.

We study the properties of vector mesons at finite temperature
with the effective lagrangian.
Here we assume that the known hadronic interactions can be extrapolated
to finite temperature and describe the interactions among particles in
hot hadronic matter.
The properties of vector mesons are modified
by the interactions with the particles in the heat bath
and this modification can be included in the self-energy.
The self-energy is defined by the difference between the inverse of the
in-medium propagator $D_{\mu\nu}$ and vacuum propagator $D^0_{\mu\nu}$
\cite{kapusta};
\begin{equation}
\Pi_{\mu\nu}=D^{-1}_{\mu\nu}-D^{-1}_{0\mu\nu}.
\end{equation}
Since the self-energy of the vector fields is the symmetric and transverse
it can be written in terms of the projection tensors as
\begin{equation}
\Pi_{\mu\nu}=GP^T_{\mu\nu}+FP^L_{\mu\nu},
\end{equation}
where $P^T_{00}=P^T_{0i}=P^T_{i0}=0$,
      $P^T_{ij}=\delta_{ij}-k_i k_j/\vec{\rm k}^2$ and
      $P^L_{\mu\nu}=k_\mu k_\nu/k^2-g_{\mu\nu}-P^T_{\mu\nu}$.
The functions $F$ and $G$ is given by
\begin{eqnarray}
F(k_0,\vec{\rm k}) &=& {k^2\over\vec{\rm k}^2}\Pi_{00}(k_0,\vec{\rm k}),\cr
G(k_0,\vec{\rm k}) &=& -{1\over2}
                        [\Pi^\mu_\mu(k_0,\vec{\rm k})+F(k_0,\vec{\rm k})],
\end{eqnarray}
where we use $k=(k_0,\vec{\rm k})$ and $k^2=k_0^2-\vec{\rm k}^2$.
The propagator of the vector mesons in the medium can be written as
\begin{equation}
{\cal D}_{\mu\nu}=-{P_L^{\mu\nu}\over k^2-m_\rho^2-F}
                  -{P_T^{\mu\nu}\over k^2-m_\rho^2-G}
                  -{k^\mu k^\nu\over m_\rho^2k^2}.
\end{equation}

We consider the corrections that come from interactions with thermal pions.
In the chiral limit, $m_\pi=0$,
the interaction with pions generates power corrections, controlled by
the expansion parameter $\sim T^2/f_\pi^2$, and the thermal corrections can be
obtained in a systematic way.
Particles with mass $M$ generate contributions of order $exp(-M/T)$ which
are exponentially suppressed compared with the effects from thermal pions.
In this respect, it is similar to calculation of the loop corrections
in chiral perturbation \cite{chpt}.
The self-energy of vector mesons in hot hadronic matter
can be expanded in powers of $T^2/f_\pi^2$ as
\begin{equation}
\Pi_{\mu\nu}(k_0,k;T)=\Pi^{(1)}_{\mu\nu}(k_0,k;T^2/f_\pi^2)
                     +\Pi^{(2)}_{\mu\nu}(k_0,k;T^4/f_\pi^4)+\cdots.
\end{equation}

The leading contributions can be obtained from one-loop diagrams in fig. 1.
Even in the presence of the $a_1$-meson propagator in the internal loop we can
still make a systematic expansion.
At the leading order in temperature the contribution from diagram fig.~1-c can
written as
\begin{equation}
\Pi^{(c)}_{\mu\nu}(k_0,\vec{\rm k})
= {1\over 2f_\pi^2} k^2(k^2 g_{\mu\nu}-k_\mu k_\nu)
T\sum\int{d^3\vec{\rm p}\over(2\pi)^3}
{1\over p^2}{1\over (p-k)^2-m_a^2}.
\end{equation}
The full expression is given in appendix.
For the axial vector mesons with momentum $q$,
we have $1/(q^2-m_a^2)$ in which
the momentum $q$ is given by the momentum of thermal pion, $p$,
and the external vector mesons, $k$. When we do the integration
over thermal pions we get
\begin{equation}
T\sum\int{d^3\vec{\rm p}\over(2\pi)^3}{1\over p^2}{1\over (p-k)^2-m_a^2}\sim
{1\over k^2-m_a^2}\int{\vert\vec{\rm p}\vert
                  d\vert\vec{\rm p}\vert\over e^{{\rm p}/T}-1}
\left(1+{\vert\vec{\rm p}\vert^2\over k^2-m_a^2}+\cdots\right),
\end{equation}
where we assume that
\begin{equation}
{\vert\vec{\rm p}\vert^2\over k^2-m_a^2}\sim
{\vert\vec{\rm p}\vert^2\over m_\rho^2} < 1
\end{equation}
With $m_\rho^2=g^2f_\pi^2$ we can still expand the correction
in powers of $T^2/f_\pi^2$.

At the leading order in $T^2$ we have
\begin{eqnarray}
F^\pi &=& {k^2\over\vec{\rm k}^2}
\left[ g^2 f_\pi^2{k_0^2\over k^2}{\epsilon\over2}
      -g^2 f_\pi^2\left(1+{\vec{\rm k}^2\over 2 m_\rho^2}\right)
       {\epsilon\over 2}\right],\cr
F^{a_1} &=& k^2\left(1+{m_a^2\over k^2-m_a^2}\right){\epsilon\over4},
\end{eqnarray}
where $F^\pi$ is the contribution comes from pion loops in figs.~1-a,b
and $F^{a_1}$ is from $\pi-a_1$ loop in fig.~1-c.
The first two terms in $F^\pi$ are the same terms obtained
from the effective model only with charged pions and neutral vector mesons
\cite{gale}. With including of the $a_1$ mesons we have $F^{a_1}$ and
the last term in $F^\pi$.
When we use the relations
\begin{equation}
g^2 f_\pi^2= m_\rho^2, \qquad m_a^2=2 m_\rho^2,
\end{equation}
there is an exact cancellation between the last term in $F^\pi$ and
the first term in $F^{a_1}$ and finally we have
\begin{equation}
F=\left(m_\rho^2+{m_\rho^4\over k^2-m_a^2}\right)\epsilon.
\end{equation}

The longitudinal mode of vector meson propagator is modified at
finite temperature and shows isospin mixing effect in the leading order
of temperature;
\begin{eqnarray}
{1\over k^2-m_\rho^2-F} &=& {1\over k^2-m_\rho^2}
+{1\over k^2-m_\rho^2}F{1\over k^2-m_\rho^2}+\cdots\cr
                        &=& (1-\epsilon){1\over k^2-m_\rho^2}
                           +\epsilon{1\over k^2-m_a^2}+ {\cal O}(T^4)
\end{eqnarray}

For transverse mode we have
\begin{equation}
G=g^2 f_\pi^2{\epsilon\over2}
  -g^2 f_\pi^2{k^2\over 2 m_\rho^2}{\epsilon\over2}
+k^2\left(1+{m_a^2\over k^2-m_a^2}\right){\epsilon\over4},
\end{equation}
which is the same as the longitudinal component.
The transverse and longitudinal components are the same at the $T^2$ order
and the mixing effect appears in both modes of vector propagator
at finite temperature.

\section{Effective masses of vector mesons}

Even though the effective lagrangian we used satisfies
the Universality and KSRF relations, the lagrangian could not reproduce the
experimental values of the masses and decay widths of $a_1$-mesons.
For given parameters with $2g\xi=1$ and $Z^2=1/2$, we have
$m_a=\sqrt{2}m_\rho=1089$ MeV and $\Gamma(a_1\to\pi\rho)=53$ MeV while
experiments show that
$m_a^{\rm exp}=1260$ MeV and $\Gamma^{\rm exp}(a_1\to\pi\rho)=400$ MeV.
By adding an extra non-minimal coupling term as \cite{sigma}
\begin{equation}
{\cal L}_\sigma=\sigma {\rm Tr} [F^L_{\mu\nu}UF^{R\mu\nu}U^\dagger].
\end{equation}
we can well describe the masses and widths of vector and axial vector
mesons with parameters; $g=10.3063,\>\sigma=0.3405,\>
\xi=0.4473$ \cite{song}.
However, the lagrangian does not satisfy Universality,
$g_{\rho\pi\pi}=3.06(g/\sqrt{2})$ and KSRF relation,
$m_\rho^2=1.63g^2f_\pi^2$.

With these parameters we have
\begin{equation}
F= g^2 f_\pi^2{\epsilon\over2}
  +k^2\left(2\eta_2^2-{\lambda g^2f_\pi^2\over m_\rho^2}\right){\epsilon\over2}
  +\eta_2^2k^2{m_a^2\over k^2-m_a^2}\epsilon,
\label{F}
\end{equation}
where $\eta_2$ and $\lambda$ are given in the appendix.
The second term in $F$ is not canceled and
the first and the last term have different coefficients.
By introducing an extra term, there is not the same mixing
in the vector and axial-vector correlator as shown from the
calculation based on current algebra and PCAC, and
the effective masses of the $\rho$-mesons have $T^2$ dependent corrections.

We obtain the effective mass from the pole position of the propagator
at zero three-momentum.
Since there is no distinction between
the longitudinal and transverse modes in the limit $\vec{\rm k}\to 0$,
the effective masses of vector mesons are obtained from the equation
\begin{equation}
k_0^2-m_\rho^2-F(k_0,\vec{\rm k}\to 0)=0.
\label{pmass}
\end{equation}
{}From eq.~(\ref{F}) and (\ref{pmass}) we see that
the effective $\rho$-meson masses are not changed with temperature
for $\sigma=0$, $2g\xi=1$ and $Z^2=1/2$.
For $\sigma\ne 0$,
we show the temperature dependence of the $\rho$ meson masses in fig. 2.
The dashed line is the result obtained from the caluction only with pions
and rho-mesons,
and the solid line is that obtained with including $a_1$-mesons
in the effective lagrangian.
The effective masses of vector mesons are still not changed much
with temperature at $T^2$ order \cite{error}.
The increase due to the pion loops is almost canceled out when
we include $a_1$ mesons.

\section{Conclusion}

We study the properties of vector mesons at finite temperature
with an effective chiral lagrangian in which the vector mesons
are introduced as massive Yang-Mills fields.
It is shown that, at the leading order of temperature,
the isospin mixing effect in vector and axial-vector correlator take places
and the effective masses of vector mesons are not changed.
When we include an extra term in the lagrangian, ${\cal L}_\sigma$,
to fit the experimental values for $a_1$ meson mass and width,
there is not the same mixing in the vector and axial-vector
correlator as shown from model independent calculation,
and there is $T^2$ dependent
correction to $\rho$ meson mass.
However, the effective masses of vector mesons are still
not changed much with temperature.
The increase of the effective masses due to
the pion loop corrections are almost canceled by the contribution from
$\pi-a_1$ loop.

This result implies that
the effect due to the modification of vector meson masses
cannot be observed unless the temperature of the hadronic matter is
very close to the critical temperature for the phase transition.
Instead, at low temperature, there is an appreciable reduction in the
coupling constant of the external vector current to vector mesons
because of the mixing effect \cite{slee}.
The reduction in the coupling constant leads to a suppression in the production
rates of photons and dileptons from hot hadronic matter,
which has stronger dependence on the temperature than the shift
of the peak position in the spectrum \cite{song2}.

\acknowledgements

The author thanks to Su Houng Lee and Che Ming Ko for useful conversations.
This work was supported by the National Science Foundation under
Grant No. PHY-9212209.

\appendix
\section{$\rho$-meson self-energy from one-loop diagrams}
The expressions for the $\rho$-meson self-energy is given by
\begin{eqnarray}
\Pi^{(a)}_{\mu\nu}(k_0,k) &=& -{1\over 2} g^2
                               T\sum\int{d^3\vec{\rm p}\over(2\pi)^3}
                               {(2p_\mu-k_\mu)(2p_\nu-k_\nu)\over p^2(p-k)^2}
                               \\[2pt]
\Pi^{(b)}_{\mu\nu}(k_0,k) &=& \left[ g_{\mu\nu}
                     -{\lambda\over m_\rho^2}(k^2g_{\mu\nu}-k_\mu k_\nu)\right]
                   g^2 T\sum\int{d^3\vec{\rm p}\over(2\pi)^3}{1\over
p^2}\\[2pt]
\Pi^{(c)}_{\mu\nu}(k_0,k) &=& {2\over f_\pi^2}
                              T\sum\int{d^3\vec{\rm p}\over(2\pi)^3}
                           {1\over p^2}{1\over (p-k)^2-m_a^2}\cr
               && \times\Biggl\{ \eta_2^2 k^2(k^2 g_{\mu\nu}-k_\mu k_\nu)
               +2\eta_2\bar\eta(k\cdot p)(k^2 g_{\mu\nu}-k_\mu k_\nu)\cr
 &&\quad+\bar\eta^2[(k\cdot p)^2g_{\mu\nu}-(k\cdot p)(p_\mu k_\nu+p_\nu k_\mu)
   +k^2p_\mu p_\nu]\cr
&&\quad +{\eta_1^2\over m_a^2}
[(k^2)^2p_\mu p_\nu-k^2(k\cdot p)
(p_\mu k_\nu+p_\nu k_\mu)+(k\cdot p)^2k_\mu k_\nu]\Biggr\}
\end{eqnarray}
where the superscript a,b,c denote the contributions from fig. 1-a,b,c,
respectively, and the
\begin{eqnarray}
\eta_1 &=& {g^2f_\pi^2\over 2m_\rho^2}
           \left({1-\sigma\over 1+\sigma}\right)^{1/2}
          +{2g\xi\over\sqrt{1+\sigma}}
           \left(1-\sigma\over 1+\sigma\right){m_\rho^2\over m_a^2}
           \\[2pt]
\eta_2 &=& {g^2f_\pi^2\over 2m_\rho^2}
           \left({1+\sigma\over 1-\sigma}\right)^{1/2}
          -{2\sigma\over\sqrt{1-\sigma^2}}\\[2pt]
\lambda &=&{g^2f_\pi^2\over 2m_\rho^2}
           \left({1+\sigma\over 1-\sigma}\right)
          -{4\sigma\over 1-\sigma}\left(1-{m_\rho^2\over g^2f_\pi^2}\right)
\end{eqnarray}
and $\bar\eta=\eta_1-\eta_2$.
When $\sigma=0$, $2g\xi=1$ and $Z^2=1/2$
we have $\eta_2=1/2$ and $\lambda=1/2$.

\begin{figure}
\caption{ One loop diagrams for the $\rho$-meson self-energy.
          The dotted, solid and double solid lines
          denote, respectively, the pion, $\rho$-meson and $a_1$-meson.}
\label{diagram}
\end{figure}

\begin{figure}
\caption{ Effective mass of $\rho$-meson at finite temperature.
          The dashed line is the result from the calculation with pions and
          rho-mesons. The solid line is the result obtained when
          $a_1$-mesons are included.}
\label{mass}
\end{figure}


\begin{references}
\bibitem{error} There are errors in the previous calculation \cite{song}.
                The coefficient $B_1$ of eq.~(35) in the ref.~\cite{song}
                should be
                $B_1=2-2k_0^2\lambda/m_\rho^2-2\eta_2\bar\eta k_0^2$
                with $\lambda$, $\eta_2$ and
                $\bar\eta$ are given in the appendix.
                With wrong coefficient the increase of $\rho$ mass at finite
                temperature was overestimated.
\end{references}
\end{document}